\documentclass[11pt]{article}
\usepackage{epsfig}
\usepackage{fullpage}

\def\captionstyle{}
\def\boxcaptionstyle{\raggedright}

\def\maxfigfraction{.6}

\newdimen\figboxmargin
\newdimen\figboxhang
\def\DVIscaling{1}
\def\globalscaling{1}
\def\figuredirectory{.}
\let\boxer=\llboxer

\def\missingfigure#1{\hbox{Missing figure #1.ps}}

\catcode `\@=11

\newbox\figurebox

\def\figbox{
\@ifnextchar[{\figboxaux}{\figboxaux[htb]}}

\long\def\figboxaux[#1#2]#3#4#5#6{
\writepict{{#3}{#4}{#5}{#6}}
\setbox\figurebox\hbox{#3}%
\if l#1\tryleftbox{#4}{#5}{#6}%
\else
\if r#1\tryrightbox{#4}{#5}{#6}%
\else
\if *#1\checktwocoloptions#2]{\box\figurebox}{#4*}{#5}{#6}%
\else\tryonecol[#1#2]{#4}{#5}{#6}%
\fi
\fi
\fi\ignorespaces}

\long\def\tryleftbox#1#2#3{
\ifdim\wd\figurebox>\maxfigfraction\columnwidth \tryonecol[htb]{#1}{#2}{#3}%
\else\leftbox{\captionbox{\box\figurebox}{#1}{#2}{#3}}\fi}

\long\def\tryrightbox#1#2#3{
\ifdim\wd\figurebox>\maxfigfraction\columnwidth \tryonecol[htb]{#1}{#2}{#3}%
\else\rightbox{\captionbox{\box\figurebox}{#1}{#2}{#3}}\fi}

\def\checktwocoloptions{
\@ifnextchar]{\floatbox[htb}{\floatbox[}}

\long\def\tryonecol[#1]#2#3#4{
\ifdim\wd\figurebox>\columnwidth \floatbox[#1]{\box\figurebox}{#2*}{#3}{#4}%
\else\floatbox[#1]{\box\figurebox}{#2}{#3}{#4}\fi}

\long\def\floatbox[#1]#2#3#4#5{%
\begin{#3}[#1]
\hbox to \hsize{\hfil#2\hfil}
\captionandlabel{#3}{#4}{#5}
\end{#3}
}

\long\def\captionbox#1#2#3#4{
\setbox\figurebox\hbox{#1}%
\parbox[t]{\wd\figurebox}{%
\bigskip\box\figurebox
\let\captionstyle=\boxcaptionstyle
\captionandlabel{#2}{#3}{#4}
\bigskip
}}

\def\captionandlabel#1#2#3{
\def\testit{#3}%
\ifx\testit\empty\else
\writecapt{{#1}{#2}{#3}}
\captypeunstarred#1*.
\getcaption#3\endc@ption
\def\testit{#2}
\ifx\testit\empty\else\label{#2}\fi
\fi}

\def\captypeunstarred#1*#2.{
\def\@captype{#1}}

\def\getcaption{\@ifnextchar[{\getcaptwo}{\getcapone}}
\long\def\getcapone#1\endc@ption{\caption[#1]%
{\def\baselinestretch{1}\Large\normalsize\captionstyle\ignorespaces #1}}
\long\def\getcaptwo[#1]#2\endc@ption{\caption[#1]%
{\def\baselinestretch{1}\Large\normalsize\captionstyle\ignorespaces #2}}

\newdimen\figboxht
\newcount\figboxn

\newcount\figboxlines
\newdimen\figboxwid

\newif\ifisleftbox

\long\def\leftbox#1{%
\setbox\figurebox\hbox{#1}\global\isleftboxtrue
\startmarginbox
\vadjust{\smash{\rlap{\hskip\hsize\hskip\figboxhang
\llap{\raise.7\baselineskip\box\figurebox\hskip\rightskip}}}}%
\endmarginbox%
}

\long\def\rightbox#1{%
\setbox\figurebox\hbox{#1}\global\isleftboxfalse
\startmarginbox
\smash{\llap{\raise.7\baselineskip\box\figurebox\hskip\figboxmargin}}%
\endmarginbox%
}

\def\startmarginbox{%
\ifvmode\passpict\let\endmarginbox=\indent
\else\message{WARNING: marginbox in not in vmode}\hfilneg\ \passpict
\let\endmarginbox=\relax\fi
\figboxht=\dp\figurebox
\advance\figboxht by 1.3\baselineskip
\vskip.95\figboxht\penalty-300\vskip-.95\figboxht
\divide\figboxht by\baselineskip
\global\figboxlines=\figboxht
\global\figboxwid=\wd\figurebox
\global\advance\figboxwid by \figboxmargin
\global\advance\figboxwid by -\figboxhang
\setmypar\noindent}
\def\addlines#1{\global\advance\figboxlines by #1\myparshape}
\def\zerolines{\origpar\global\figboxlines=0\myparshape}

\def\passpict{\par\ifnum\figboxlines>1\vskip\figboxlines\baselineskip
\zerolines\fi}

\def\emptybox#1#2{\hbox to #1{\vbox to #2{\vss}\hss}}

\global\let\origpar=\@@par
\global\let\dopar=\origpar
\global\def\@@par{\dopar}
\@setpar{\dopar}

\def\setmypar{\global\let\dopar=\mypar
\global\prevgraf=0\myparshape}

\def\mypar{\origpar\global\advance\figboxlines by -\prevgraf%
\global\prevgraf=0\myparshape}

\def\myparshape{\relax%
\ifnum\figboxlines>1\theparshape \else
\global\hangindent=0pt\global\hangafter=1
\global\let\dopar=\origpar\fi}

\def\theparshape{%
\ifisleftbox\global\hangindent=-\figboxwid 
\else\global\hangindent=\figboxwid \fi
\global\hangafter=-\figboxlines \global\advance\hangafter by 1%
}

\def\definefnum#1{
\def\fnum@figure{Figure \ref{#1}}%
\def\fnum@table{Table \ref{#1}}%
\def\fnum@code{Algorithm \ref{#1}}%
}

\def\writepict#1{}
\def\writecapt#1{}

\def\journalpicts#1{
\newwrite\pictfile
\newwrite\captfile
\openout\pictfile\jobname.pic
\openout\captfile\jobname.cap
\gdef\writepict##1{\unexpandedwrite\pictfile{\doit##1}}%
\gdef\writecapt##1{\unexpandedwrite\captfile{\doit##1}}%
\global\let\ENDdocument=\enddocument
\gdef\enddocument{\DOjournalpicts{#1}\ENDdocument}
}

\def\DOjournalpicts#1{{%
\def\writepict##1{}\closeout\pictfile
\def\writecapt##1{}\closeout\captfile
\@fileswfalse
\onecolumn
\def\globalscaling{#1}
\def\doit##1##2##3##4{
\figboxaux[t]{\hss##1\hss}{##2}{}{}%
\vspace*{1in}
\definefnum{##3}
\captionandlabel{##2}{}{##4}
\clearpage}%
\input main2.pic
\def\doit##1##2##3{
\definefnum{##2}
\captionandlabel{##1}{}{##3}}%
\raggedright\let\captionstyle=\raggedright
\def\@makecaption##1##2{##1: ##2\par}
\input\jobname.cap
}}

\def\llboxer#1{\vbox to \figboxht{\vfil\hbox to \figboxwid{#1\hfill}}}
\def\lcboxer#1{\vbox to \figboxht{\vfil\hbox to \figboxwid{\hfill#1\hfill}}}
\def\oldboxer#1{\vbox to \figboxht{\vfil
                      \hbox to \figboxwid{\hfill\llap{#1\hskip4.25in}\hfill}}}
\def\ulboxer#1{\vbox to \figboxht{\hbox to \figboxwid{#1\hfill}\vfil}}
\def\ccboxer#1{\vbox to \figboxht{\vfil
                        \hbox to \figboxwid{\hfill#1\hfill}\vfil}}
\def\basicboxer#1{\vbox to \figboxht{\vfil\hbox to \figboxwid{#1}\vskip-.1in}}

{\catcode`\p=12\catcode`\t=12
\gdef\removedimen#1pt{#1}}

\def\defscaled#1#2{#2=\DVIscaling#2%
\xdef#1{\expandafter\removedimen\the#2}}

\def\DVIspace{ }
\newdimen\hscalefactor
\newdimen\vscalefactor

\def\scale#1{\horizscale{#1}\vertscale{#1}}
\def\horizscale#1{\hscalefactor=#1\hscalefactor\figboxht=#1\figboxht}
\def\vertscale#1{\vscalefactor=#1\vscalefactor\figboxwid=#1\figboxwid}

\def\boxps{%
\@ifnextchar[{\boxpsaux}{\boxpsaux[\relax]}}

\def\boxpsaux[#1]#2#3#4#5{%
{\figboxwid#4\figboxht#5\hscalefactor=1pt\vscalefactor=1pt%
\scale{#3}%
\scale{\globalscaling}%
#1%
\defscaled\DVIhscale\hscalefactor%
\defscaled\DVIvscale\vscalefactor%
\boxer{\includegraphics{\figuredirectory/#2.ps}}}%
}

\newread\Epsffilein
\newif\ifEpsffileok
\newif\ifEpsfbbfound

\newdimen\pspoints
\divide\pspoints by 72

\def\boxeps{%
\@ifnextchar[{\boxepsaux}{\boxepsaux[\relax]}}
\def\boxepsaux[#1]#2#3{%
\openin\Epsffilein=\figuredirectory/#2.ps 
\ifeof\Epsffilein\message{I couldn't open \figuredirectory/#2.ps }%
\missingfigure{#2}
\else
   {\Epsffileoktrue\Epsfbbfoundfalse
    \catcode`\%=11 \catcode`\\=11
    \catcode`\{=11 \catcode`\}=11
    \catcode`\$=11 \catcode`\^=11
    \catcode`\&=11 \catcode`\#=11
    \catcode`\~=11 \catcode`\_=11
    \loop
       \read\Epsffilein to \Epsffileline
       \ifeof\Epsffilein\Epsffileokfalse\else
          \expandafter\Epsfaux\Epsffileline . .\\%
       \fi
   \ifEpsffileok\repeat
   \ifEpsfbbfound
        \figboxht=\Epsfury\pspoints
        \advance\figboxht by-\Epsflly\pspoints
        \figboxwid=\Epsfurx\pspoints
        \advance\figboxwid by-\Epsfllx\pspoints
   \else
        \message{No bounding box comment in \figuredirectory/#2.ps }%
        \figboxwid=2in\figboxht=1in%
   \fi%
   \immediate\closein\Epsffilein
   \hscalefactor=1pt\vscalefactor=1pt%
   \scale{#3}%
   \scale{\globalscaling}%
   #1%
   \defscaled\DVIhscale\hscalefactor
   \defscaled\DVIvscale\vscalefactor
   \hscalefactor=-\Epsfllx\hscalefactor
   \hscalefactor=1.00375\hscalefactor
   \defscaled\DVIhoffset\hscalefactor
   \vscalefactor=-\Epsflly\vscalefactor
   \vscalefactor=1.00375\vscalefactor
   \defscaled\DVIvoffset\vscalefactor
   \llboxer{\includegraphics{\figuredirectory/#2.ps}} }%
\fi
}%
{\catcode`\%=11 \global\let\Epsfpar=\par
\global\let\Epsfpercent=
\long\def\Epsfaux#1#2 #3\\{\relax\ifx#1\Epsfpercent
   \def\testit{#2}\ifx\testit\Epsfbblit
      \Epsfsize #3 . . . .\\%
      \global\Epsffileokfalse
      \global\Epsfbbfoundtrue
   \fi\else\ifx#1\Epsfpar\else\global\Epsffileokfalse\fi\fi}%
\def\Epsfsize#1 #2 #3 #4 #5\\{\global\def\Epsfllx{#1}\global\def\Epsflly{#2}%
   \global\def\Epsfurx{#3}\global\def\Epsfury{#4}}%

\catcode`\@=12                  

\def\pic#1;#2;#3;#4\par{\picsc#1;#2;#3;1;#4\par}

\def\picsc#1;#2;#3;#4;#5\par{
\figbox[htb]{\boxeps{#1}{#4}
}{figure}{#1}{%
#5}}

\def\mpic#1;#2;#3;#4\par{\mpicsc#1;#2;#3;1;#4\par}

\def\mpicsc#1;#2;#3;#4;#5\par{
\figbox[l]{\boxeps{#1}{#4}
}{figure}{#1}{%
#5}}

\bibdata{bibdata}
\bibliographystyle{alpha}

\begin{document}


\nocite{*}


\newtheorem{theorem}{Theorem}
\newcommand{\stephLine}{\rule{\linewidth}{2mm}}

\newenvironment{stephTitle}
{\fontencoding{OT1}\fontfamily{cmss}\fontseries{bx}\fontshape{n}%
\fontsize{36}{48pt}\selectfont}{}

\newenvironment{stephSubTitle}
{\fontencoding{OT1}\fontfamily{cmss}\fontseries{bx}\fontshape{n}%
\fontsize{12}{12pt}\selectfont}{}


\author{Alex Brodsky\thanks{Research supported by NSERC PGSB}, 
Stephane Durocher, 
Ellen Gethner\thanks{{\tt \{abrodsky,durocher,egethner\}@cs.ubc.ca},
Department of Computer Science, 
University of British Columbia, 201 - 2366 Main Mall,
Vancouver, B.C., Canada,
V6T 1Z4}}
\title{Toward the Rectilinear Crossing Number of $K_n$: New Drawings,
Upper Bounds, and Asymptotics}
\date{September 14, 2000}
\maketitle
\begin{abstract}
Scheinerman and Wilf \cite{wilf} assert that ``an important open 
problem in the study of graph
embeddings is to determine the rectilinear crossing number of the complete 
graph $K_n$.''  A 
{\it rectilinear drawing of $K_n$} is an arrangement of $n$ 
vertices in the plane, every
pair of which is connected by an edge that
is a line segment. We assume that no three vertices are collinear, and that 
no three edges intersect in a point unless that point is an endpoint of all
three.
The {\it rectilinear crossing number of $K_n$} is 
the fewest number of edge crossings attainable over all rectilinear
drawings of $K_n$.

For each $n$ we construct a rectilinear drawing of $K_n$ that has
the fewest number of edge crossings and the best asymptotics known to date. 
Moreover, we give some
alternative infinite families of drawings of $K_n$ with good asymptotics. 
Finally, we mention some 
old and new open problems.
\end{abstract}

{\bf keywords} crossing number, rectilinear, complete graph

\section{Introduction and History}

Given an arbitrary graph $G$, determining a drawing of $G$ in the plane
that produces the fewest number of edge crossings
is NP-Complete \cite{GJ83}. The complexity is not known for an arbitrary 
graph when the edges are assumed
to be line segments \cite{bienstock}. 
Recent exciting work on the general crossing 
number problem (where edges are
simply homeomorphs of the unit interval
$[0,1]$ rather than line segments) has been
accomplished by Pach, Spencer, and T{\'o}th~\cite{pach}, who give a tight 
lower bound for the crossing number of families of  graphs with certain forbidden 
subgraphs. We study 
the specific instance of determining the {\bf rectilinear crossing number of 
$K_n$}, denoted {\bf $\overline{cr}(K_n)$}, and 
we offer 
drawings with ``few'' edge crossings. 
The difficulty of determining the exact value of $\overline{cr}(K_n)$, even 
for small values of $n$, manifests
itself in the sparsity
of literature \cite{guy2,erdos,singer,bdg3}. Other contributions
are given as general 
constructions \cite{jensen,hayward} that yield
upper bounds and asymptotics, none of which lead to exact 
values of $\overline{cr}(K_n)$ for all $n$. Finally, there is an elegant
and surprising connection between the asymptotics of the rectilinear crossing
number of $K_n$ and Sylvester's four point problem of geometric probability
\cite{wilf, wilf2}.

Much of the information regarding progress of any kind has been 
disseminated by personal communication, and now in this era of ``the 
information 
highway,'' some revealing sources of the unfolding story can be found on the
web \cite{finch, archdeacon}. 

In this paper we offer new constructions, upper bounds, and asymptotics, 
which we motivate and
explain by the interesting and nondeterministic historical progress of the 
problem and its elusive solution.

\section{Recursive Construction of $K_n$}
\subsection{Introduction}
\figbox[l]{\boxeps{concentric}{0.4}}{figure}{fig:concentric}{concentric versus 
non-concentric triangles}
\noindent Upon examining different configurations of vertices in the plane,
one quickly realizes that drawings that minimize crossings tend to
have vertices aligned along three axes, forming a triangular 
structure of nested concentric triangles; such configurations are 
``opposite'' in flavour to placing vertices on a convex hull.
Two nested triangles $t_1$ and $t_2$ are {\bf concentric} if and only if 
any edge with endpoints in $t_1$ and $t_2$ does not intersect any edge
of $t_1$ or $t_2$ (see Figure~\ref{fig:concentric}).
In $K_4$ through $K_9$, for which optimal drawings are known~\cite{guy2,white}, 
the tripartite pattern is evident.
The same pattern exists in generalized constructions presented by 
Jensen \cite{jensen}
and Hayward \cite{hayward} for any $K_n$.

\figbox[r]{\boxeps{JenHay}{0.7}}{figure}{fig:JenHay}{positioning vertices 
using Jensen's \cite{jensen}
and Hayward's \cite{hayward}
constructions}
Various schemes are possible for positioning vertices within
each of the three parts. 
In Jensen's construction, vertices along an axis are positioned by 
alternating above and below the axis (see Figure~\ref{fig:JenHay}a). 
In Hayward's construction, vertices along an axis are 
are positioned on a concave curve (see Figure~\ref{fig:JenHay}b).
Alternatively, 
the collection of vertices along each axis could be 
arranged 
to minimize crossings within the collection, while maintaining 
concentricity of the triangles.
We examine a construction and variations, originally suggested by 
Singer~\cite{singer},
that positions vertices along each axis by recursive definition of
similarly constructed smaller graphs.

\subsection{Definitions}
We identify
specific sets of edges, sets of vertices, and subgraphs, within the
larger construction of $K_n$. 
Those components of the graph that are recursively defined form
{\bf clustervertices}. Each clustervertex is itself a
complete graph $K_a$, where $a < n$; a clustervertex with $a$ vertices
is said to have {\bf order} $a$. 
If both endpoints of an edge $uw$ are
contained within clustervertex $c$, then $uw$ is {\bf internal}
to $c$. Similarly, a vertex $w$ contained within a clustervertex $c$ is 
internal to $c$.
Given two clustervertices $c_1$ and $c_2$,
the set of all edges that have one endpoint in each of $c_1$ and $c_2$
form a {\bf clusteredge}. Finally, if $q$ clusteredges meet at 
clustervertex $c$,
then $c$ has {\bf clusterdegree} $q$.

\figbox[l]{\boxeps{crush}{0.4}}{figure}{fig:crush}{flattening a clustervertex}
Recursively constructed clustervertices are {\bf flattened} by an affine 
transformation~\cite[Ch.~15]{martin}. 
Vertices appear as a sequence of nearly collinear vertices. Of course, no three
vertices in the graph can be collinear, thus the flattened clustervertex has
some height $\epsilon > 0$
(see Figure~\ref{fig:crush}) and its edge crossings are unaltered by the
scaling.
When a clustervertex $c$ is flattened, an incident clusteredge $e$ is said to
{\bf dock} at $c$. 
Given a flat clustervertex $c$, two clusteredges $e_1$ and $e_2$
may dock at $c$
from opposite sides such that no 
edge crossings are created between $e_1$ and $e_2$. 
When two clusteredges $e_1$ and $e_2$ dock on the same side of a clustervertex
$c$, we say $e_1$ and $e_2$ {\bf merge} at $c$ (see Figure~\ref{fig:HEcrossHE}).

\subsection{Counting Toolbox}
Given a generalized definition for graph construction involving
clustervertex interconnection, 
the following functions count edge crossings for 
the various types of edge intersections. 

\subsubsection{$f(k)$: Single Vertex Docked at a Clustervertex}
\figbox[l]{\boxeps{fcross3}{0.38}}{figure}{fig:crossNeigh}{Edge $uw$ crosses at
most six internal edges.}
When a new vertex $u$ is created, 
new edges are added from $u$ to all other existing vertices. 
Specifically, given a clustervertex $c$ of order $k$, 
an edge must be added from $u$ to
every vertex in $c$. An edge from $u$ to a vertex $w$ in $c$ may cross
some internal edges of $c$. If $w$ is the $i$th vertex
in the sequence of vertices of $c$, $i-1$ vertices lie on one side of 
$w$ in $c$ and $k-i$ vertices lie on the other side 
(see Figure~\ref{fig:crossNeigh}a).
Thus, edge $uw$ will be required to cross 
at most $(i-1)(k-i)$ edges of $c$. If we add edges from $u$ to every vertex
in $c$, the number of new edge crossings within $c$ will be at most
\begin{equation}
f(k) = \sum_{i=1}^k(i-1)(k-i) = \frac{k^3}{6} - \frac{k^2}{2} + \frac{k}{3} \ .
\end{equation}

If we add two vertices $v_1$ and $v_2$ on opposite sides of a clustervertex $c$,
then for every internal vertex $w$ of $c$, 
the internal edges that span $w$ will be crossed exactly once, either by
edge $v_1w$ or by edge $v_2w$ but not both (see Figure~\ref{fig:crossNeigh}b).
The number of new edge crossings among vertices of $c$ and $v_1$ and $v_2$
will be exactly $f(k)$.

\subsubsection{$i(p,k)$: Internal Clusteredge Intersections}
Given two clustervertices $c_k$ and $c_p$ of orders $k$ and $p$, and 
a clusteredge $e$ between them that docks completely on one side of each
clustervertex, selecting two vertices from each clustervertex forms a
quadrilateral that contributes one edge crossing.  The number of 
edge crossings within $e$ is given by
\begin{equation}
i(p,k) = {p \choose 2} {k \choose 2} = \frac{p(p-1)k(k-1)}{4} \ .
\end{equation}

\subsubsection{$e(k,p,j)$: Two Clusteredges Merge at a Clustervertex}
\figbox[r]{\boxeps{HEcrossHE}{0.415}}{figure}{fig:HEcrossHE}{two clusteredges 
merge at a clustervertex}
When two clusteredges originate from clustervertices of orders $p$ and $j$ and
merge at a clustervertex of order $k$
(see Figure~\ref{fig:HEcrossHE})
the number of crossings between
edges of the two clusteredges (ignoring crossings with
edges internal to the clustervertex)
is given by
\begin{equation}
e(k,p,j) = \sum_{i=0}^{k-1} ipj = \frac{pjk(k-1)}{2} \ .
\end{equation}
If the two clusteredges intersect away from a clustervertex, then the number of
crossings is simply $p \cdot j \cdot k \cdot l$, where the clusteredge crossing
is between four clustervertices of orders $p$, $j$, $k$, and $l$.

\subsection{Recursive Definitions}
The following constructions of $K_n$ involve recursive definition by 
connecting $q$ clustervertices $K_k$ of order $k$, where $n = q \cdot k$.
Scheinerman and Wilf show that $\overline{cr}(K_n) =
\Theta(n^4)$~\cite{wilf}.
In a worst case drawing, where edge crossings are maximized,
every subset of four vertices contributes one edge crossing.
This occurs when all vertices lie on a convex hull, creating ${n \choose 4}$
crossings. Thus, when a better drawing is found, we examine what
fraction of the crossings remain by taking
the limit of $g(n) / {n \choose 4}$
as $n \rightarrow \infty$, where $g(n)$ is a count of the crossings in 
the new drawing.

\subsubsection{Triangular Definition}
\figbox[l]{\boxeps{k3new}{0.39}}{figure}{fig:k3rec}
{$K_n$ defined by three $K_{n/3}$}
Singer suggests a recursive construction \cite{singer,wilf2} where,
given $n=3^j$, we draw $K_n$ by taking three flat instances of $K_{n/3}$
and adding new edges (see Figure~\ref{fig:k3rec}). Each instance of 
$K_{n/3}$ is drawn recursively. $K_3$ gives a base 
case.

Let $k = n/3$ and let $C_3(n)$ represent the total number of crossings in 
$K_n$ under the drawing defined by this recursive construction.
There are $C_3(k)$ crossings 
internal to each of the clustervertices, $k \cdot f(k)$ crossings for each
clustervertex corresponding to clusteredge to clustervertex dockings, and
$i(k,k)$ crossings internal to each clusteredge.

Given that $C_3(3) = 0$, the total number of crossings
is given by
\begin{eqnarray}
C_3(n) & = & 3C_3(k) + 3k \cdot f(k) + 3i(k,k) 
 =  \frac{5}{312}n^4 - \frac{1}{8}n^3 + \frac{7}{24}n^2 - \frac{19}{104}n
\end{eqnarray}
\begin{equation}
\Rightarrow\lim_{n \rightarrow \infty} \frac{C_3(n)}{{n \choose 4}} = \frac{15}{39} 
\approx 0.3846 \ .
\end{equation}

\subsubsection{Recursive Definitions Using a Larger $K_a$}
Just as we do for $K_3$, 
we may use any optimal drawing of $K_a$ 
as a recursive template. 
Given $n = a^j$, we apply an analogous procedure where
clustervertices are defined recursively.
In addition to counting recursive terms, $C_a(k)$, internal clusteredge
crossings, $i(k,k)$, and clusteredge-clustervertex crossings, $k\cdot f(k)$,
we must also count pairs of clusteredges that merge, $e(k,k,k)$, and
clusteredge crossings away from a clustervertex, $k^4$.
Using $K_4$ as a basis and $C_4(4)=0$, we derive
\begin{eqnarray}
C_4(n)  =  4C_4(k) + 6 i(k,k) + 6 k \cdot f(k) + 4 e(k,k,k) 
 =  \frac{1}{56}n^4 - \frac{2}{15}n^3 + \frac{7}{24}n^2 - \frac{37}{210}n
\end{eqnarray}
\begin{equation}
\Rightarrow\lim_{n \rightarrow \infty} \frac{C_4(n)}{{n \choose 4}} = \frac{3}{7} 
\approx 0.4286 \ .
\end{equation}

Using $K_5$ as a basis and $C_5(5)=1$, we derive
\begin{eqnarray}
C_5(n)=5C_5(k) + 10 i(k,k) + 10 k \cdot f(k) 
 + 10 e(k,k,k) + k^4 
= \frac{61}{3720}n^4 - \frac{1}{8}n^3 + \frac{7}{24}n^2 - \frac{227}{1240}n
\end{eqnarray}
\begin{equation}
\Rightarrow\lim_{n \rightarrow \infty} \frac{C_5(n)}{{n \choose 4}} = \frac{227}{155}
\approx 0.3935 \ .
\end{equation}

\figbox[r]{\boxeps{HEatHV}{0.3}}{figure}{fig:HEatHV}
{balanced clusteredge dockings}
Similarly, we derive limits using $K_7$ and $K_9$ as templates (see Table~1). 
As one
would expect, 
the limit for $K_9$ is equal to that for $K_3$, since both are powers 
of three.
For any odd $a$, 
we derive a generalized exact count using a recursive $K_a$ construction.
We require a count for the number of crossings in $K_a$, both for
our base case, $C_a(a) = \overline{cr}(K_a)$, and for 
recursively-defined clusteredge to clusteredge crossings.

The count breaks down as follows. Let $k = n/a$. We take 
$a$ recursive instances of $K_k$ which contribute $a \cdot C_a(k)$
crossings. We add crossings for every pair of clusteredges that merge at a 
clustervertex. Each clustervertex has clusterdegree $a-1$. To minimize
crossings, clusteredges must be split evenly on either side of a flattened
clustervertex (see Figure~\ref{fig:HEatHV}).
Thus, clusteredge dockings contribute
$2a{(a-1)/2 \choose 2} e(k,k,k)$ crossings.
Pairs of dockings on opposite sides of a clustervertex contribute
exactly ${a \choose 2} k \cdot f(k)$ crossings.
\figbox[l]{\mbox{
\begin{tabular}{|r|c|c|l|}
\hline
 & $a$ & $\lim_{n\rightarrow \infty} \frac{g(n)}{{n \choose 4}}$ & comment\\
\hline
Singer \cite{singer} & 3 & 0.3846 & $n = 3^j,$ $C_3(3) = 0$ \\
Brodsky-Durocher-Gethner & 4 & 0.4286 & $n = 4^j,$ $C_4(4) = 0$ \\
Brodsky-Durocher-Gethner & 5 & 0.3935 & $n = 5^j,$ $C_5(5) = 1$ \\
Brodsky-Durocher-Gethner & 7 & 0.3885 & $n = 7^j,$ $C_7(7) = 9$ \\
Brodsky-Durocher-Gethner & 9 & 0.3846 & $n = 9^j,$ $C_9(9) = 36$ \\
\hline
Jensen \cite{jensen} & -- & 0.3888 & any $n$ \\
Hayward \cite{hayward} & -- & 0.4074 & any $n$ \\
Scheinerman-Wilf \cite{wilf} & -- & 0.2905 & lower bound \\
Guy \cite{guy} & -- & 0.3750 & conjectured $cr(K_n)$ (non-rectilinear) \\
\hline
\end{tabular}      
}}{table}{table1}{asymptotics for $C_a(n)$ compared with known bounds}
Clusteredges have internal crossings that add another
${a \choose 2} i(k,k)$. 
Finally, we must account for clusteredge to clusteredge crossings that
occur in $K_a$ itself; thus we add $\overline{cr}(K_a) \cdot k^4$.
This gives
\begin{eqnarray}
C_a(n) = 
	a \cdot C_a(k) + {a \choose 2} k \cdot f(k) 
	+ 2a{\frac{a-1}{2} \choose 2} e(k,k,k) + {a \choose 2} i(k,k)
	+ \overline{cr}(K_a) \cdot k^4 \ .
\end{eqnarray}

We can solve for a non-recursive closed form of $C_a(n)$ by simplifying
\begin{equation}
\frac{n}{a} \overline{cr}(K_a)
	+ \sum_{j=1}^{\log_an-1} a^{j-1} \left[
	  {a \choose 2} k \cdot f(k) + {a \choose 2} i(k,k) 
	+ 2a{\frac{a-1}{2} \choose 2} e(k,k,k) 
	+ \overline{cr}(K_a) \cdot k^4 \right], 
\end{equation}
where $k = n/a^j$. 

Out of all recursive constructions for which $\overline{cr}(K_a)$ is known, 
the best results are achieved by
$C_3(n)$ (see Table~1). 
The construction can easily be generalized by dividing $n$ into
three parts of sizes $\lfloor\frac{n}{3}\rfloor$, $\lceil\frac{n}{3}\rceil$, 
and $n - \lfloor\frac{n}{3}\rfloor - \lceil\frac{n}{3}\rceil$. Since two of
the three parts will always have the same size, $f(k)$ always gives an
exact count. 
By induction,
one can show that $C_{3g}(n) < jen(n)$ for $n\geq 24$, where $C_{3g}(n)$ is
a count of the crossings in the generalized 
construction and $jen(n)$ is
the number of crossings in $K_n$ using Jensen's construction\footnote{$jen(n)=\left\lfloor\frac{7n^4-56n^3+128n^2+48n\lfloor \frac{n-7}{3}\rfloor +108}{432} \right\rfloor$}\cite{jensen}.
Thus, asymptotically, \mbox{$C_{3g} < 3\cdot [jen(k) + k\cdot f(k) + i(k,k)]$}, 
with $k = n/3$, 
and we get an
upper bound of 0.3848 for a general $n$. In the next section we offer some improvements.

\section{Asymptotic Improvements}
Within the recursive constructions presented thus far,
edges arriving at a flattened clustervertex are balanced; 
if $q$ edges arrive at clustervertex $c$ of degree $p$, 
then exactly $q/2$ edges arrive
at $c$ from each side and $\frac{q}{2}f(p)$ crossings are added. 
However, depending on the side of entry,
the number of edges crossed when entering a
clustervertex differs. Thus, it may be advantageous to have an imbalance
in the number of edges docking on each side of a clustervertex. 

Most of the crossings in $C_3(n)$ occur at the top level of the recurrence,
as is shown by
\begin{equation}
\lim_{n\rightarrow \infty} \frac{C_3(n) - 3C_3(n/3)}{C_3(n)} = \frac{26}{27}\ .
\end{equation}
\figbox[r]{\boxeps{slide}{0.35}}{figure}{fig:slide}{sliding a clustervertex}
Improving the top level of the construction while 
slightly compromising on recursive constructions could reduce the total
crossings. Improvements at the top-level can be achieved
by moving clustervertices to alter the number of edges that
reach a neighbouring clustervertex from above and from below
(see Figure~\ref{fig:slide}). In doing so, however, new crossings
are created at the merging of clusteredges. Thus, there exists
a point of balance that minimizes total\\ 
\indent\indent\indent\indent\indent\indent\indent\indent\indent\indent~~~~~~~
crossings lost and gained
by the translation.

\subsection{Maximally Asymmetric Internal Clustervertices}
\figbox[l]{\boxeps{arc}{0.85}}{figure}{fig:arc}{minimizing crossings from above}
In the extreme case, we construct each of the three partitions by taking
a convex $K_k$ (see Figures~\ref{fig:arc} and~\ref{fig:slideShape}). 
Crossings from
above are minimized and crossings from below are maximized to form
a maximally asymmetric drawing.

Let $k = n/3$ and let $a+b=k$ determine how much to slide the clustervertex, 
where $b$ is a measure of how many vertices in one clustervertex change position
relative to the other two. Assuming each
clustervertex is moved by the same amount, the top-level graph will appear
as in Figure~\ref{fig:slideShape}. Accounting using the usual
tools gives the following count of crossings
{\small
\begin{eqnarray}
C_m(n,a) &=& 3 \left[{k \choose 4} + a\cdot f(k) + i(a,a) 
 + i(b,b) + 2i(a,b) \right. \nonumber \\
&& \left. + e(a,b,b) + 2e(a,a,b) 
 + e(b,b,b) + 2e(b,a,b) + ab^3 + a^2b^2 \right]
\nonumber \\
&=& \frac{19}{648}n^4-\frac{5}{54}n^3a+\frac{1}{6}n^2a^2 
- \frac{5}{36}n^3 + \frac{1}{6}n^2a - \frac{1}{2}na^2 + \frac{17}{72}n^2
+ \frac{1}{3}na - \frac{1}{4}n.
\end{eqnarray}}

$C_m(n,a)$ is a  quadratic polynomial in $a$ and
is minimized when $a_0=5n/18+1/3$. This gives
{\small
\begin{equation}
C_m(n,a_0) = \frac{4}{243}n^4 - \frac{85}{648}n^3 + \frac{67}{216}n^2 
- \frac{7}{36}n 
\end{equation}
\begin{equation}
\Rightarrow\lim_{n \rightarrow \infty} \frac{C_m(n,a_0)}{{n \choose 4}} 
= \frac{32}{81} \approx 0.3951 \ . 
\end{equation}}
$C_3(n)$ still performs better than $C_m(n,a)$ for any $a$. Thus, using 
convex $K_k$ as
first-level clustervertices overcompensates the savings of the
recursive structure in $C_3(n)$. Therefore, we define a new construction
that maintains the 
recursive structure of $C_3(n)$ for clustervertices.

\subsection{Retaining $C_3(n)$ as Internal Clustervertices}
\figbox[r]{\boxeps{k3topbot}{0.35}}{figure}{fig:k3topbot}{docking above versus
below}
Previously, $f(k)$ counted access into an internal clustervertex $c$ of
order $k$,
where dockings were balanced on both sides of $c$.
For imbalanced access, we derive a separate count of edge crossings entering
$c$ from above and
from below where $c$ is recursively defined by $C_3(k)$ and $k=3^j$.
In the base cases, $n=3$, no crossings occur above and a single crossing
occurs below. Thus, we define $f_{top}(3) = 0$ and $f_{bot}(3) = 1$.
Assume the triangles are arranged recursively to point upwards. We count 
crossings as follows. Assume $k = n/3$.
If the new point is positioned above the clustervertex, $3\cdot f_{top}(k)$
edges are crossed recursively and $3\cdot e(k,k,1)$ are crossed at the
top-level. If the new point is positioned below the clustervertex, then
$k^3$ additional crossings occur (see Figure~\ref{fig:k3topbot}). 
Thus, we derive the following recurrences:
\begin{eqnarray}
f_{top}(n)  =  3 [f_{top}(k) + e(k,k,1)] 
= \frac{n^3}{16} - \frac{n^2}{4} + \frac{3n}{16} \\
f_{bot}(n)  =  3 [f_{bot}(k) + e(k,k,1)] + k^3 
= \frac{5n^3}{48} - \frac{n^2}{4} + \frac{7n}{48} \ .
\end{eqnarray}

As expected, $f(n) = f_{top}(n) + f_{bot}(n)$. The difference
between $f_{top}(n)$ and $f_{bot}(n)$ is significant as is shown by,
\begin{equation}
\lim_{n\rightarrow \infty} \frac{f_{top}(n)}{f_{bot}(n)} = \frac{3}{5}
\end{equation}

\figbox[l]{\boxeps{globSlide3}{0.35}}{figure}{fig:slideShape}{Clustervertices 
are not actually broken, only
translated; they are drawn as two parts for counting.
Clusteredges are drawn as arcs to reduce clutter.}
Sliding a clustervertex creates new crossings at the merging of
two clusteredges and at the crossing of new clusteredges 
(see Figure~\ref{fig:slideShape}b). We count the cost of sliding
one, two, or three clustervertices. These counts are given by $C_{s1}(n,a)$,
$C_{s2}(n,a)$, and $C_{s3}(n,a)$, respectively.
For each, $a$ represents the portion
of the affected clustervertex that still docks on the same side of incident
clustervertices. 
$a$ is defined in terms of $n$. When more than
one clustervertex is moved, both or all three being moved are moved by the
same amount. 
\figbox[l]{\mbox{
\begin{tabular}{|r|c|c|c|c|c|}
\hline
&graph 		& internal 	& top-level 	& total      & minimizing $a_0$\\
\hline
Singer\cite{singer} &$C_3(n)$& 0.0142 	& 0.3704 	& 0.3846 
& \\
Brodsky-Durocher-Gethner&$C_m(n,a)$& 0.0370	& 0.3580& 0.3951
&$a_0=5n/18+1/3$\\
Brodsky-Durocher-Gethner&$C_{s1}(n,a)$ & 0.0142& 0.3701	& 0.3843   
& $a_0=23n/72 - 1/24$\\
Brodsky-Durocher-Gethner&$C_{s2}(n,a)$ & 0.0142& 0.3699        & 0.3841     
& $a_0=23n/72 - 1/24$\\
Brodsky-Durocher-Gethner&$C_{s3}(n,a)$ & 0.0142& 0.3696        & 0.3838    
& $a_0=23n/72 - 1/24$\\
\hline
\end{tabular}
}}{table}{table2}{asymptotic improvements on $C_3(n)$}

Using a counting argument identical to that for $C_m(n,a)$, we derive the
following:
{\small
\begin{eqnarray}
C_{s1}(n,a) &=& \frac{137}{6318}n^4 - \frac{23}{648}n^3a + \frac{1}{18}n^2a^2
- \frac{31}{216}n^3 + \frac{1}{9}n^2a - \frac{1}{6}na^2 + \frac{8}{27}n^2
- \frac{1}{72}na - \frac{19}{104}n,\\
C_{s2}(n,a) &=& \frac{691}{25272}n^4 - \frac{23}{324}n^3a + \frac{1}{9}n^2a^2
- \frac{35}{216}n^3 + \frac{2}{9}n^2a - \frac{1}{3}na^2 + \frac{65}{216}n^2
- \frac{1}{36}na - \frac{19}{104}n,\\
C_{s3}(n,a) &=& \frac{139}{4212}n^4 - \frac{23}{216}n^3a + \frac{1}{6}n^2a^2
- \frac{13}{72}n^3 + \frac{1}{3}n^2a - \frac{1}{2}na^2 + \frac{11}{36}n^2
- \frac{1}{24}na - \frac{19}{104}n.
\end{eqnarray}  
}

Again, each count is quadratic with respect to $a$ and 
each is minimized when $a_0 = 23n/72 - 1/24$.  The value $a$ 
represents the number of vertices in a clustervertex that dock on the
bottom of the clustervertex on its (counter-clockwise) right side.
Thus, we require $a$ to be an integer.
One observes, however, that $a_0 = 23n/72-1/24$ is never an integer for $n=3^i$, but
an induction argument shows that $\lceil 23n/72-1/24\rceil$ is the integer
nearest $a_0$. 
Let $a_1(j) = 3^j\cdot 23/72-1/24$ and 
let $a_2(j) = \lceil3^j\cdot 23/72-1/24 \rceil$. 
Asymptotically, $C_{s3}(n,a)$ remains unaffected since
\begin{equation}
\forall \epsilon >0,\ \exists i \in \mathbf{Z}\ s.t.\ \forall j>i\  
\left| \frac{C_{s3}(3^j,a_1(j))}{{3^j \choose 4}} 
- \frac{C_{s3}(3^j,a_2(j))}{{3^j \choose 4}} \right| < \epsilon\ . 
\end{equation}
To obtain the number of edge crossings for a given $n=3^i$ and
$a_0 = 23n/72 - 1/24$, simply evaluate $ C_{s3}(n,\lceil a_0 \rceil)$. Thus
\begin{equation}
\overline{cr}(K_n)\leq  C_{s3}(n,\lceil 23n/72-1/24 \rceil)
\ .
\end{equation}
Asymptotically, this value approaches $C_{s3}(n, a_0)$, which gives
\begin{equation}
C_{s3}(n,a_0) = \frac{6467}{404352}n^4 - \frac{1297}{10368}n^3
+ \frac{1009}{3456}n^2 - \frac{2723}{14976}n\ .
\end{equation}
A similar argument holds for $C_{s1}(n,a)$ and $C_{s2}(n,a)$.
Thus, we derive the following limits:
\begin{eqnarray}
\lim_{n \rightarrow \infty} \frac{C_{s1}(n,a_0)}{{n \choose 4}} 
&=& \frac{19427}{50544} \approx 0.3846, \\
\lim_{n \rightarrow \infty} \frac{C_{s2}(n,a_0)}{{n \choose 4}}
&=& \frac{9707}{25272} \approx 0.3841, \\
\lim_{n \rightarrow \infty} \frac{C_{s3}(n,a_0)}{{n \choose 4}}
&=& \frac{6467}{16848} \approx 0.3838 \ .
\end{eqnarray}

\subsection{Generalized Upper Bounds}

\begin{theorem}\label{thm:main}
\begin{equation}
\lim_{n\rightarrow \infty} \frac{\overline{cr}(K_n)}{{n \choose 4}} \leq 
\frac{6467}{16848} \approx 0.3838
\end{equation}
\end{theorem}

\noindent {\bf Proof.} 
Scheinerman and Wilf show that $\overline{cr}(K_n)/ {n \choose 4}$
is a nondecreasing function \cite{wilf}. We know 
$ \overline{cr}(K_n) \leq C_{s3}(n,a_0)$ for all $n=3^i$. 
Therefore, 
\begin{equation}\label{asymptotics}
\lim_{n\rightarrow \infty} \frac{\overline{cr}(K_n)}{{n \choose 4}}
\leq \lim_{n\rightarrow \infty} \frac{C_{s3}(n,a_0)}{{n \choose 4}}
= \frac{6467}{16848}\ .\  \rule{5pt}{5pt}
\end{equation}

As we did for $C_3(n)$, our construction for $C_{s3}(n,a)$ can be generalized 
by dividing $n$ into three partitions of sizes $p_1$, $p_2$, and $p_3$
such that $\max_{i,j} | p_i-p_j| \leq 1$. Each partition then forms a 
clustervertex
defined recursively by $C_{3g}(p_i)$. Clustervertices are translated by
an appropriate $a_i$ that is the integer nearest $23p_i/72 - 1/24$.
We conjecture that such
constructions produce asymptotics close to those achieved in
Theorem~\ref{thm:main}.

We also mention recent work on a new lower bound in equation
(\ref{asymptotics}) based on work accomplished in \cite{bdg3}. That is,
$\overline{cr}(K_{10})=62$, from which it follows that $.3001\leq 
\lim_{n \rightarrow \infty} \frac{C_{s3}(n,a_0)}{{n \choose 4}}$. In summary we have

\begin{equation}
.3001\leq 
 \lim_{n \rightarrow \infty}\frac{C_{s3}(n,a_0)}{{n \choose 4}}\leq .3838.
\end{equation}

\subsection{Example: $K_{81}$}
\begin{figure}[h]
\begin{center}
\ \mbox{\epsfig{file=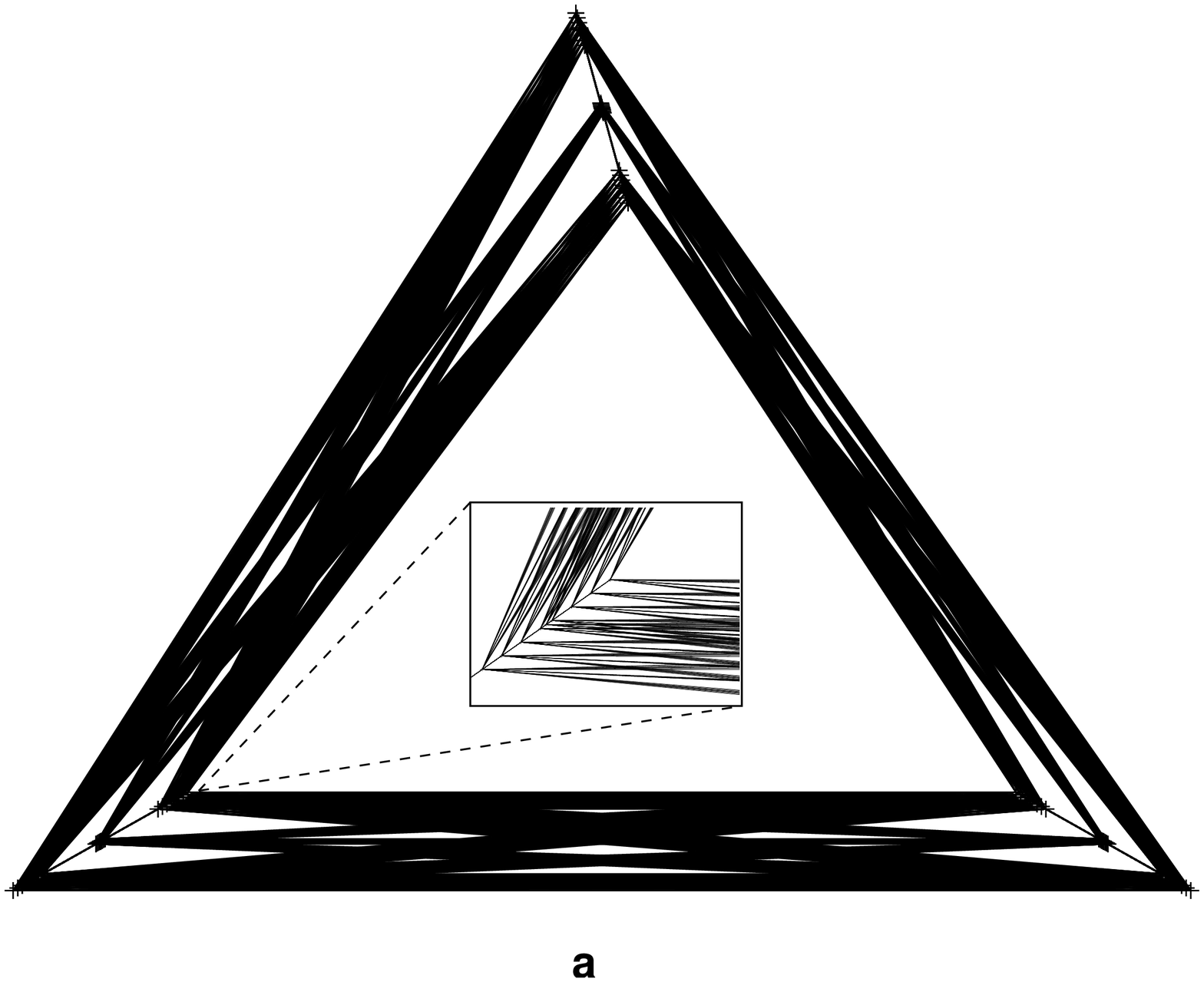,width=0.4\linewidth}
        \epsfig{file=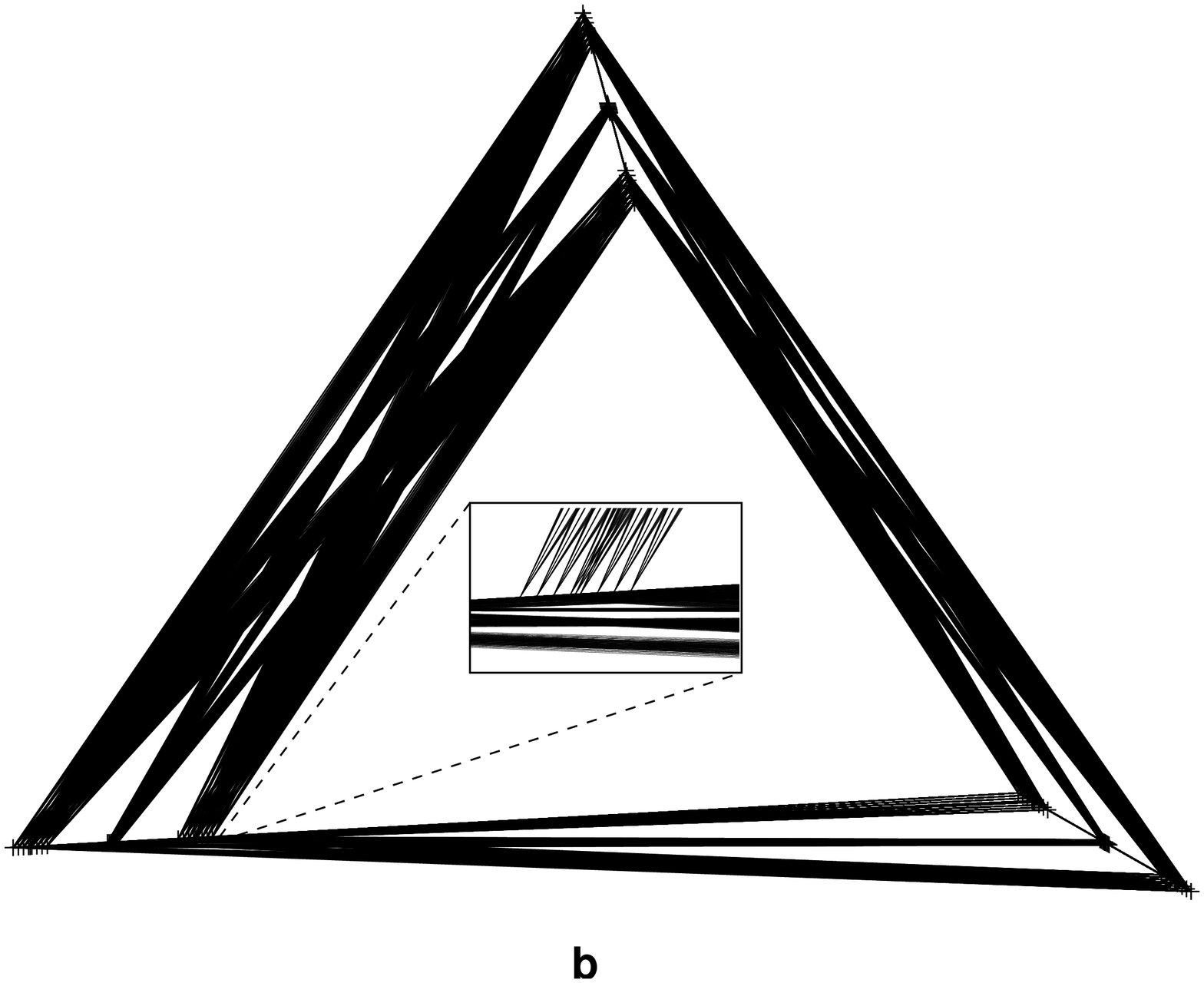,width=0.4\linewidth}}\ 
\end{center}
\caption{two instances of $K_{81}$\label{fig:oldvsnew}}
\end{figure}

In Figure~\ref{fig:oldvsnew}, we give two rectilinear drawings of
$K_{81}$. The first drawing is based on Singer's
construction~\cite{singer,wilf2} and has $625,320$ edge crossings. The second
drawing\footnote{These calculations were verified by an arbitrary
precision edge-crossing counter.} is based on the construction given
by the strategy corresponding to $C_{s1}(81,26)=624,852$.

\begin{table}[h]
\begin{center}
\begin{tabular}{|l|c|c|c|c|c|c|c|c|}
\hline
strategy & $C_{s3}(81,26)$ & $C_{s2}(81,26)$ & $C_{s1}(81,26)$ & 
          $C_3(81)$\cite{singer} & \cite{jensen} & \cite{hayward}&  ${81 \choose 4}$ \\ 
\hline
count    & $623,916$ & $624,384$ & $624,852$ & $625,320$ & $630,786$ & $659,178$ & 1,663,740   \\
\hline
\end{tabular}
\end{center}
\caption{drawings of $K_{81}$ that count}
\end{table}

The largest number of edge crossings in a rectilinear drawing of 
$K_{81}$ is ${81 \choose 4} = 1,663,740$ and occurs when all 81 vertices 
are placed on a convex hull. The fewest number of edge crossings of $K_{81}$ known
to date is $C_{s3}(81,26)=623,916$.

\section{Summary and Future Work}
In summary, most forward progress toward determining
$\overline{cr}(K_n)$ has been accomplished by producing a good
rectilinear drawing of $K_n$ for each $n$. A  ``good'' rectilinear
drawing of $K_n$ has relatively few edge crossings and avails
itself of an exact count of said crossings.  Throughout the history of
the problem, drawings that have produced the best asymptotic results
amount to iteratively producing three clustervertices, which upon
examination of the whole graph, yield a configuration of nested
concentric triangles. Our best closed form and asymptotics arose from a
break in tradition by yielding a graph with three clustervertices
forming a set of nested triangles, but whose triangles are {\it not}
pairwise concentric. 

We offer the following open question: can one extend the
technique given in Section~3 to produce a graph with more than three
clustervertices that will yield better upper bounds and asymptotics for
$\overline{cr}(K_n)$?  Singer's rectilinear drawing of $K_{10}$ with
62 edge crossings \cite{gardner, singer} was the first successful
recorded instance of this break with tradition. Additionally, can the
technique given in Section~3  be applied successfully to other families
of interesting graphs? See, for example, the work of Bienstock and Dean~\cite{dean1,dean2}. 

Our second open question is based on the current rapidly changing
status of computing, which makes feasible the use of brute-force techniques
in extracting information about small graphs. In particular, it is possible
to determine the exact value of  $\overline{cr}(K_n)$ for small values
of $n$ beyond what is presently known \cite{guy2,white,bdg3}. For example, 
a complete catalogue of inequivalent drawings is available through $n=6$
for both rectilinear and non-rectilinear drawings of $K_n$
\cite{harborth1, harborth2}. As the catalogue grows, exact values
for $\overline{cr}(K_n)$ will be found. The catalogue is being extended
computationally
by Applegate,
Dash, Dean, and Cook \cite{dean}. Additionally,  Harris and Thorpe
\cite{thorpe} have accomplished
a randomized search and produced drawings of $K_{12}$ and $K_{13}$ with 155
and 229 edge crossings respectively. Both drawings have fewer edge crossings
than the drawings given by Jensen \cite{jensen}. Our question is the
following: how many inequivalent
drawings of $K_n$ produce a number of edge crossings equal to
$\overline{cr}(K_n)$? Experimental work leads us to believe that
the answer to this question is nontrivial. As more concrete information
becomes available,
we will
be better able to investigate this question. Lastly, we note that 
Brodsky, Durocher,
and Gethner \cite{bdg3} have given a combinatorial proof that
$\overline{cr}(K_{10})=62$. We know of only one drawing of $K_{10}$ with
62 edge crossings. 

Our third and final open question concerns a
problem addressed by Hayward~\cite{hayward} and Newborn and Moser
\cite{newborn} and is the following: find a rectilinear drawing of
$K_n$ that produces the largest possible number of crossing-free
Hamiltonian cycles.
Hayward,
building on the work in \cite{newborn} has asymptotics based on
a generalized rectilinear drawing of $K_n$, as mentioned in
Section~3, Table~1.  Our construction given in Section~3 improves
Hawyard's result \cite{bdg2}. A related open problem is: does some
rectilinear drawing of $K_n$ with the minimum number of edge crossings
necessarily produce the optimal number of crossing-free Hamiltonian cycles? 
Hayward
conjectures that the answer is ``yes,'' as do we, but as of yet, no
proof is known.

Crossing number problems are rich and numerous
with much work to be done. For an excellent exposition of further
diverse open questions, see \cite{pach2}.


\section{Acknowledgments}
The authors wish to thank David Kirkpatrick and Nick Pippenger for stimulating discussions and ideas. We are also
grateful to David Singer who provided us with his manuscript \cite{singer}. We extend special thanks
to Mike Albertson, Heidi Burgiel, and Imrich Vrto who provided invaluable help with our literature search. Finally, we are indebted to Richard Guy who generously shared his
knowledge and insight with us.

\bibliography{bibdata}

\end{document}